\documentclass[12pt]{article}
\usepackage{graphics}
\usepackage{bm}
\usepackage{graphicx}
\usepackage{amsmath}
\usepackage{amssymb}
\usepackage{amstext}
\usepackage{epstopdf}
\usepackage{amscd}
\usepackage{amsfonts}
\usepackage{appendix}
\usepackage{color}
\usepackage{hyperref}
\usepackage{textcomp}
\DeclareMathAlphabet{\EuFrak}{U}{euf}{m}{n}
\DeclareMathAlphabet{\EuScript}{U}{eus}{m}{n}

\newcommand{\nd}{\noindent}

\newcommand{\be}{\begin{equation}}
\newcommand{\ee}{\end{equation}}
\newcommand{\ben}{\begin{eqnarray}}
\newcommand{\een}{\end{eqnarray}}

\title{{{\bf Newton's gravitation-force's classical average 
proof of  a  Verlinde's conjecture}}}
\author{{\small{A. Plastino$^{1,3,4}$, M. C. Rocca$^{1,2,3}$}}, \\
\small{$^1$ Departamento de F\'{\i}sica,
Universidad Nacional de La Plata,}\\
\small{$^2$ Departamento de Matem\'{a}tica,
Universidad Nacional de La Plata,}\\
\small{$^3$ Consejo Nacional de Investigaciones Cient\'{\i}ficas
y Tecnol\'{o}gicas}\\
\small{(IFLP-CCT-CONICET)-C. C. 727, 1900 La Plata -
Argentina}\\
\small{$^4$  SThAR - EPFL, Lausanne, Switzerland}}

\date{\today}

\begin{document}

\maketitle

\begin{abstract}

\nd
A surprising,  gravity related Verlinde-conjecture, that generated immense interest,  asserts that gravity is
an emergent entropic force. We provided a classical proof of the assertion in [doi.org/j.physa.2018.03.019]. 
Here, we classically prove a related, second Verlinde-conjecture. This  states that,  
 at very large distances ($r_0$),
gravity departs from its classical   nature
and begins to decay linearly with $r_0$.\\ 
\nd KEYWORDS: Verlinde's second conjecture,
Boltzmann-Gibbs distribution, divergences,  
dimensional regularization.

\end{abstract}

\renewcommand{\theequation}{\arabic{section}.\arabic{equation}}

\section{Introduction}

\nd Entropic gravity, or emergent one, is a Verlinde's conjecture that has attracted immense interest 
\cite{verlinde}.  He describes gravity as an entropic force (EF). This EF is endowed with macro-scale homogeneity, being at the same time  subjected to quantum-disorder. Accordingly, this two-body EF would not be a fundamental interaction.  Verlinde based his ideas on string theory, black hole physics, and quantum information theory. We provided a classical proof of the assertion in \cite{1}. \vskip 3mm
 \nd Such EF theory involves what we call here a second conjecture. It says that when gravity becomes vanishingly weak, at very large distances, it differs from its Newtonian quadratic  nature because its strength starts to decay linearly with the inverse distance from a given mass. We intend to prove below this second conjecture.
\vskip 3mm
\nd The main technical ingredient of our proceedings is dimensional regularization (DR). DR \cite{tq1,tp1} constitutes one of the
greatest advances in the theoretical physics of the
last 45 years, with applications in several branches
of physics (see, for instance, \cite{dr1}-\cite{dr54}. 

\vskip 3mm \nd It was believed that the classical Boltzmann-Gibbs (BG) 
probability distribution can not yield finite results because the associated partition function 
${\cal Z}$ diverges \cite{lb,grav1}.  This belief did not take into account the
possibility of  analytical extensions, that could overcome divergences, e.g., at the origin. 
However, it was shown in Refs.\cite{z,z1},  that
${\cal Z}$ can be calculated and yields finite results  for Boltzmann-Gibbs and Tsallis entropies, using the 45-years old DR technique.
\vskip 3mm \nd It is well known that, {\it at a quantum field theory level},  DR can not cope with the gravitational field, since it is non-renormalizable. Our present challenge is quite different, though, because we deal with Newton's gravity at a {\it classical} level and we are not attempting  renormalization.

\vskip 3mm
\nd We prove the second conjecture in three dimensions in Section 2 and in two dimensions in Section 3. 
The ensuing conclusions are drawn in Section 4.

\setcounter{equation}{0}

\section{The three-dimensional case}

\nd In \cite{1},  we classically verified Verlinde's emergent gravitation conjecture by starting with the ideal gas Hamiltonian, constructing the associated partition function, and from it the entropy. Then, following Verlinde's prescription for an entropic force, we showed that it had Newton's appearance. Now instead, we start from the gravitation Hamiltonian and compute the concomitant partition function. \vskip 3mm 

\nd   The Boltzmann-Gibbs (BG) partition function ${\cal Z}_\nu$ for a Newton potential
 $\frac {GmM} {r}$ is \cite{z1}
\begin{equation}
\label{ep2.1}
{\cal Z}_\nu=\int\limits_{\cal W} e^{-\beta\left(\frac {p^2} {2m}-
\frac {GmM} {r}\right)}d^\nu xd^\nu p,
\end{equation}
where the masses involved are $M$ and $m$. We call ${\cal W}= R^\nu\oplus S^\nu(r_0)$, $S^\nu(r_0)$ being the
spherical volume of radius $r_0$.
 For effecting the integration process one uses hyper-spherical coordinates and two integrals, each in  $\nu$ dimensions. 
One is then left with just two radial coordinates (one in $r-$ space and the other in $p-$ space) and 
$2(\nu -1)$ angles. Accordingly \cite{z1}, 
\begin{equation}
\label{ep2.2}
{\cal Z}_\nu=
\left[\frac {2\pi^{\frac {\nu} {2}}} {\Gamma\left(
\frac {\nu} {2}\right)}\right]^2
\int\limits_0^{\infty}
e^{-\beta\frac {p^2} {2m}}
p^{\nu-1}dp
\int\limits_0^{r_0}
e^{\beta\frac {GmM} {r}}
r^{\nu-1}dr.
\end{equation}
\vskip 3mm \nd We appeal here to a Table if Integrals \cite{gr} the integral
\begin{equation}
\label{ep2.3}
\int\limits_0^{\infty}
e^{-\beta\frac {p^2} {2m}}
p^{\nu-1}dp=
\left(\frac {2m} {\beta}\right)^{\frac {\nu} {2}}
\frac {\Gamma\left(\frac {\nu} {2}\right)} {2}.
\end{equation}
\vskip 3mm \nd The remaining integral is cast as 
\[\int\limits_0^{r_0}
e^{\beta\frac {GmM} {r}}
r^{\nu-1}dr=
\int\limits_0^{\infty}
e^{\beta\frac {GmM} {r}}
r^{\nu-1}dr-
\int\limits_{r_0}^{\infty}
e^{\beta\frac {GmM} {r}}
r^{\nu-1}dr=\]
\begin{equation}
\label{ep2.4}
\Gamma(-\nu)\cos\pi\nu(\beta GmM)^{\nu}+
\frac {r_0^\nu} {\nu}\phi\left(-\nu, 1-\nu; \frac {\beta GmM} {r_0}\right).
\end{equation}
The first integral on the r.h.s has been evaluated in \cite{z1} while the second can be read  off
 \cite{gr}. We call  $\phi$ the confluent hypergeometric function.
 We arrive in this way to the following expression for ${\cal Z}_\nu$
\begin{equation}
\label{ep2.5}
{\cal Z}_\nu=
\frac {2} {\Gamma\left(\frac {\nu} {2}\right)}
\left(\frac {2\pi^2m} {\beta}\right)^{\frac {\nu} {2}}\left[
\Gamma(-\nu)\cos\pi\nu(\beta GmM)^{\nu}+
\frac {r_0^\nu} {\nu}\phi\left(-\nu, 1-\nu; \frac {\beta GmM} {r_0}\right)\right].
\end{equation}
\vskip 3mm \nd We need further appeal to  \cite{gr}  to find
\[\phi(-\nu,1-\nu; z)=e^z\phi(1, 1-\nu; -z)=
e^z\left[1+\frac {z} {\nu-1}+\frac {z^2} {(\nu-1)(\nu-2)}+\right.\]
\begin{equation}
\label{ep2.6}
\left.\frac {z^3} {(\nu-1)(\nu-2)(\nu-3)}\phi(1, 4-\nu; -z)\right],
\end{equation}
refining further ${\cal Z}$ as
\[{\cal Z}_\nu=\frac {2} {\Gamma\left(\frac {\nu} {2}\right)}\cos\pi\nu
\left(2\pi^2\beta G^2m^3M^2\right)^{\frac {\nu} {2}}
\Gamma(-\nu)+\]
\[\frac {2} {\Gamma\left(\frac {\nu} {2}\right)}
\left(\frac {2\pi^2m} {\beta}\right)^{\frac {\nu} {2}}\frac {r_0^\nu} {\nu}
e^{\frac {\beta GmM} {r_0}}\left[1+\frac {\beta GmM} {(\nu-1)r_0}+
\frac {(\beta GmM)^2} {(\nu-1)(\nu-2)r_0^2}\right]+\]
\begin{equation}
\label{ep2.7}
\frac {2} {\Gamma\left(\frac {\nu} {2}\right)}
\left(\frac {2\pi^2m} {\beta}\right)^{\frac {\nu} {2}}
\frac {r_0^{\nu-3}} {\nu}
e^{\frac {\beta GmM} {r_0}}
\frac {(\beta GmM)^3} {(\nu-1)(\nu-2)(\nu-3)}
\phi\left(1, 4-\nu; -\frac {\beta GmM} {r_0}\right).
\end{equation}
\vskip 3mm \nd From (\ref{ep2.7}) one gathers that poles emerge  for any dimension $\nu$, 
 $\nu=3$ included. Thus, appeal to dimensional regularization (DR) is mandatory. To this effect we  use the   DR-Bollini @ Giambiagi's technique's generalization given in \cite{tp1}.   
In a nut-shell the (DR) process consists in this procedure: 
if we have, for instance, an expression $F(\nu)$ that diverges, say, for $\nu=3$, our
 Bollini-Giambiagi's DR generalization consists in performing the 
Laurent-expansion of $F$ around $\nu=3$ and select
afterwards, as the physical result for $F$, the $\nu=3$-independent term in the
expansion. The justification for such a procedure is clearly explained in \cite{tp1}.\vskip 3mm

\nd The   ${\cal Z}_\nu$'s Laurent expansion reads
\begin{equation}
\label{ep2.8}
{\cal Z}_\nu=\frac {a_{-1}} {\nu-3}+ a_0+\sum\limits_{n=0}^{\infty}
a_n(\nu-3)^n. 
\end{equation} 
Physically then, from ${\cal Z}$, DR selects  the $a_0$-term, i.e., 
\[{\cal Z}=a_0=-\frac {1} {3\sqrt{\pi}}(2\pi^2\beta G^2m^3M^2)^{\frac {3} {2}}
\left[\ln(2\pi^2\beta G^2m^3M^2)-\boldsymbol{C}-\frac {17} {3}\right]+\]
\[\frac {4} {3\sqrt{\pi}}\left(\frac {2\pi^2m} {\beta}\right)^{\frac {3} {2}}
r_0^3e^{\frac {\beta GmM} {r_0}}\left[1+\frac {\beta GmM} {2r_0}
+\frac {(\beta GmM)^2} {2r_0^2}\right]+\] 
\[\frac {(2\pi^2\beta G^2m^3M^2)^{\frac {3} {2}}} {3\sqrt{\pi}}
\left[\ln\left(\frac {2\pi^2mr_0^2} {\beta}\right)+\boldsymbol{C}+
2\ln 2-\frac {17} {3}\right]-\]
\begin{equation}
\label{ep2.9} 
\frac {2} {3\sqrt{\pi}}e^{\frac {\beta GmM} {r_0}}(2\pi^2\beta G^2m^3M^2)^{\frac {3} {2}}
\phi^{(1)}\left.\left(1, 4-\nu; -\frac {\beta GmM} {r_0}\right)\right]_{\nu=3}.
\end{equation}
where $\phi^{(1)}$ denotes derivative of $\phi$ with respect to $4-\nu$ \cite{jmp}.
\vskip 3mm \nd 
We now analyze the 4 lines that make up Eq. (\ref{ep2.9}) for 
very large $r_0$. In such an instance, the first line is constant, the second line grows as $r_0^3$, 
the third line grows logarithmically,  and the fourth one is constant.
Thus, when we pass to the limit of very large $r_0$ we find
\begin{equation}
\label{ep2.10} 
{\cal Z}=\frac {4} {3\sqrt{\pi}}\left(\frac {2\pi^2m} {\beta}\right)^{\frac {3} {2}}r_0^3.
\end{equation}
\vskip 3mm \nd We consider next the mean energy $<U>$ and face
\begin{equation}
\label{ep2.11}
<{\cal U}>_\nu=\frac {1} {{\cal Z}}
\int\limits_{\cal M}e^{-\beta\left(\frac {p^2} {2m}-
\frac {GmM} {r}\right)}
\left(\frac {p^2} {2m}-\frac {GmM} {r}\right)d^\nu xd^\nu p.
\end{equation}
\vskip 3mm \nd We have to treat $<U>$  now in identical manner as we did above with ${\cal Z}$. We do not give the pertinent details to save space. The ensuing result reads 
\[<{\cal U}>_\nu=-\frac {\nu} {\beta\Gamma\left(\frac {\nu} {2}\right)}\cos\pi\nu
\left(2\pi^2\beta G^2m^3M^2\right)^{\frac {\nu} {2}}
\Gamma(-\nu)+\]
\[\frac {1} {\beta\Gamma\left(\frac {\nu} {2}\right)}
\left(\frac {2\pi^2m} {\beta}\right)^{\frac {\nu} {2}}r_0^\nu
e^{\frac {\beta GmM} {r_0}}\left[1-\frac {\beta GmM} {(\nu-1)r_0}-
\frac {(\beta GmM)^2} {(\nu-1)(\nu-2)r_0^2}\right]-\]
\begin{equation}
\label{ep2.12}
\frac {1} {\beta\Gamma\left(\frac {\nu} {2}\right)}
\left(\frac {2\pi^2m} {\beta}\right)^{\frac {\nu} {2}}
r_0^{\nu-3}
e^{\frac {\beta GmM} {r_0}}
\frac {(\beta GmM)^3} {(\nu-1)(\nu-2)(\nu-3)}
\phi\left(1, 4-\nu; -\frac {\beta GmM} {r_0}\right).
\end{equation} \vskip 3mm \nd 
Again,  poles in $\nu$ ensue and we need DR once again, that is,
 the Laurent series for $<{\cal U}>_\nu$ around $\nu=3$. We arrive at 
\begin{equation}
\label{ep2.13}
<{\cal U}>_\nu=\frac {1} {{\cal Z}}\left[\frac {b_{-1}} {\nu-3}+ b_0+\sum\limits_{n=0}^{\infty}
b_n(\nu-3)^n\right]. 
\end{equation} 
and the physical term for ${\cal U}$, the one independent of $\nu-3$, is now 
\[<{\cal U}>=\frac {b_0} {{\cal Z}}=\frac {1} {{\cal Z}}\left\{
\frac {1} {2\beta\sqrt{\pi}}(2\pi^2\beta G^2m^3M^2)^{\frac {3} {2}}
\left[\ln(2\pi^2\beta G^2m^3M^2)-\boldsymbol{C}-5\right]\right.+\]
\[\frac {2} {\beta\sqrt{\pi}}\left(\frac {2\pi^2m} {\beta}\right)^{\frac {3} {2}}
r_0^3e^{\frac {\beta GmM} {r_0}}\left[1-\frac {\beta GmM} {2r_0}
-\frac {(\beta GmM)^2} {2r_0^2}\right]+\] 
\[\frac {(2\pi^2\beta G^2m^3M^2)^3} {2\beta\sqrt{\pi}}
\left[\ln\left(\frac {8\pi^2mr_0^2} {\beta}\right)+\boldsymbol{C}-5\right]-\]
\begin{equation}
\label{ep2.14} 
\left.\frac {1} {\beta\sqrt{\pi}}e^{\frac {\beta GmM} {r_0}}
(2\pi^2\beta G^2m^3M^2)^{\frac {3} {2}}
\phi^{(1)}\left.\left(1, 4-\nu; -\frac {\beta GmM} {r_0}\right)\right]_{\nu=3}\right\}.
\end{equation}  \vskip 3mm \nd 
Proceeding similarly to what was done with ${\cal Z}$ we have for very large 
$r_0$:
\begin{equation}
\label{ep2.15}
<{\cal U}>=\frac {3} {2\beta}.
\end{equation}
The following abbreviation is useful: 
\begin{equation}
\label{ep2.16} 
{\cal Z}=\alpha r_0^3\;\;\;\;\;;\;\;\;\;\;
\alpha=\frac {4} {3\sqrt{\pi}}\left(\frac {2\pi^2m} {\beta}\right)^{\frac {3} {2}}.
\end{equation} \vskip 3mm \nd 
The entropy in the canonical ensemble reads now \cite{z1}
\begin{equation}
\label{ep2.17} 
{\cal S}=\ln{\cal Z}+\beta<{\cal U}>=\ln{\cal Z}+\frac {3} {2}
\end{equation}
for very large  $r_0$. Verlinde's entropic force is defined as 
\begin{equation}
\label{ep2.18} 
F_e=-\frac{\lambda} {\beta}\frac {\partial{\cal S}} {\partial r_0}=
-\frac {\lambda} {{\beta\cal Z}}\frac {\partial{\cal Z}} {\partial r_0},
\end{equation}
i.e., 
\begin{equation}
\label{ep2.19} 
F_e=-\frac{3\lambda} {\beta r_0}=
-\frac{GmM} { r_0},
\end{equation}
with $\lambda$
\begin{equation}
\label{ep2.20} 
\lambda=
\frac{\beta GmM} {3}.
\end{equation} \vskip 3mm \nd 
We realize that  (\ref{ep2.19}) does have for $F_e$ the form, at large distance,  
conjectured by  Verlinde, QED. This is our main conclusion in the present Communication.

\setcounter{equation}{0}

\section{The planar case}

In two dimensions one faces

\begin{equation}
\label{ep3.1}
{\cal Z}_2=\int\limits_{\cal M} e^{-\beta\left(\frac {p^2} {2m}+
2GmM\ln r\right)}d^2xd^2p,
\end{equation}
where ${\cal M}=R^2\oplus S^2(r_0)$ and $S^2(r_0)$ is the
spherical volume of radius $r_0$.   Using polar coordinates one has 
\begin{equation}
\label{ep3.2}
{\cal Z}_\nu=
4\pi^2
\int\limits_0^{\infty}
e^{-\beta\frac {p^2} {2m}}
pdp
\int\limits_0^{r_0}
e^{-2\beta GmM\ln r}
rdr
\end{equation}
or

\begin{equation}
\label{ep3.3}
{\cal Z}=
4\pi^2
\int\limits_0^{\infty}
e^{-\beta\frac {p^2} {2m}}
pdp
\int\limits_0^{r_0}
r^{1-2\beta GmM}dr.
\end{equation}
The first integral on the r.h.s. of  (\ref{ep3.3}) is straightforward. 
The second needs appeal to the integral regularization technique of I. M. Guelfand in Vol. 1
of his treatise  \cite{IMG}. This leads to

\begin{equation}
\label{ep3.4}
\int\limits_0^{r_0}
r^{1-2\beta GmM}dr=\frac {r_0^{2-2\beta GmM}} {2-2\beta GmM}
\;\;\;\;\;;\;\;\;\;\;2-2\beta GmM\neq 0.
\end{equation}
The planar partition function becomes
\begin{equation}
\label{ep3.5}
{\cal Z}=\frac {4\pi^2m} {\beta}\frac {r_0^{2-2\beta GmM}} {2-2\beta GmM},
\end{equation}
while $<U>$ is

\begin{equation}
\label{ep3.6}
<{\cal U}>=\frac {1} {{\cal Z}}
\int\limits_Me^{-\beta\left(\frac {p^2} {2m}+
2Gmm\ln r\right)}
\left(\frac {p^2} {2m}+2GmM\ln r\right)d^\nu xd^\nu p.
\end{equation} 
We follow the steps of the preceding Section and obtain for the partition function times $<U>$

\begin{equation}
\label{ep3.7}
{\cal Z}<{\cal U}>=\frac {4\pi^2m} {\beta^2}\frac {r_0^{2-2\beta GmM}} {2-2\beta GmM}\left[
1+2\beta GmM\left(\ln r_0-\frac {1} {2-2\beta GmM}\right)\right],
\end{equation}
or
\begin{equation}
\label{ep3.8}
<{\cal U}>=\frac {1} {\beta}\left[
1+2\beta GmM\left(\ln r_0-\frac {1} {2-2\beta GmM}\right)\right].
\end{equation}
From large $r_0$ one finds, from (\ref{ep3.8}):
\begin{equation}
\label{ep3.9}
<{\cal U}>=\frac {1} {\beta}\left(
1+2\beta GmM\ln r_0\right).
\end{equation}
Passing to the entropic force we face
\begin{equation}
\label{ep3.10} 
F_e=-\frac{2\lambda} {\beta r_0}=
-\frac{GmM} { r_0},
\end{equation}
with 
\begin{equation}
\label{ep3.11} 
\lambda=
\frac{\beta GmM} {2}.
\end{equation}
The statistically averaged planar entropic force's behaviour coincides with that of  
 the three-dimensional one at very large $r$. This fact might tempt one to conjecture that, at large distances, the mass-distribution should be planar. 

\section{Conclusions}

\nd Two inspiring Verlinde's conjectures regarding the  gravitational interaction  have been proved at a classical statistical level. \vskip 3mm
 \nd First, that it is an emergent force derived from entropy, proved in \cite{1}. Second, proved here, that at at very large distances the interaction decays as $1/r$ and not as $1/r^2$. \vskip 3mm

\nd Verlinde has revolutionized our conception of gravity. Here we have contributed our grain of sand to such revolution.

\setcounter{equation}{0}

\nd

\end{document}